\documentclass[prl,showpacs,preprintnumbers,amsmath,amssymb,tightenlines,twocolumn]{revtex4}

\usepackage{graphicx}% Include figure files
\usepackage{dcolumn}% Align table columns on decimal point
\usepackage{bm}% bold math
\usepackage{epsfig}
\usepackage{stmaryrd}
\usepackage{color}

\newcommand{\bra}[1]{\langle\,{#1}\, |}
\newcommand{\ket}[1]{|\,{#1}\,\rangle}
\newcommand{\braket}[2]{\mbox{$\langle\,{#1}\, | \,{#2}\,\rangle$}}

\newcommand{\HamTot}{H}
\newcommand{\HamEl}{H^{\rm el}}
\newcommand{\Vdd}{V}

\newcommand{\Ort}{R}

\newcommand{\Pos}{\Ort}
\newcommand{\PosAbs}{\Ort}

\newcommand{\atomA}{2}
\newcommand{\atomB}{3}
\newcommand{\pdiff}[2]{\frac{\partial #1}{\partial #2}}

\newcommand{\sub}[2]{{#1}_{\mbox{\!\! \scriptsize #2}}}

\newcommand{\bv}[1]{\mathbf{ #1 }}

\def\beq{\begin{equation}}
\def\eeq{\end{equation}}

\def\CR{\nonumber\\[0.15cm]}

\newcommand{\fref}[1]{Fig.~\ref{#1}}
\newcommand{\frefp}[2]{Fig.~\ref{#1}~(#2)}

\newcommand{\eref}[1]{Eq.~(\ref{#1})}

\newcommand{\cref}[1]{chapter~\ref{#1}}

\newcommand{\Cref}[1]{Chapter~\ref{#1}}

\newcommand{\bref}[1]{(\ref{#1})}

\usepackage{ulem}  %unter- (\uline{important}) und durchstreichen (\sout{wrong})
\begin{document}

\title{Newton's cradle and entanglement transport in a flexible Rydberg chain}
\author{S.~W\"uster}
\author{C.~Ates}
\author{A.~Eisfeld}
\author{J.~M.~Rost}
\affiliation{Max Planck Institute for the Physics of Complex Systems, N\"othnitzer Strasse 38, 01187 Dresden, Germany}
\email{sew654@pks.mpg.de}
\begin{abstract}
In a regular, flexible chain of Rydberg atoms, a single electronic excitation localizes on two atoms that are in closer mutual proximity than all others. 
We show how the interplay between excitonic and atomic motion causes electronic excitation and diatomic proximity to propagate through the Rydberg chain as a combined pulse.
In this manner entanglement is transferred adiabatically along the chain, reminiscent of momentum transfer in Newton's cradle. 
\end{abstract}
\pacs{
32.80.Ee,  % Rydberg States
82.20.Rp,  % State to state energy transfer 
34.20.Cf    % Interatomic potentials and forces
}
\maketitle

%%%%%%%%%%%%%%%%%%%%%%%%%%%%%%%%%%%
%%%%%%%%%%%%%%%%%%%%%%%%%%%%%%%%%%%

Nearly loss-less transfer of momentum and energy through a linear chain of masses was studied as early as in the 17$^{\rm th}$ century, 
exemplified by Newton's cradle~\cite{herrmann:cradle}. Newton's cradle, the well known table-top experiment of popular physics, in which the first and last member of a suspended chain of metal balls periodically exchange oscillation energy, has recently found a modern implementation with ultra-cold atoms \cite{kinoshita:cradle}.
Stimulated by research on photosynthesis~\cite{AmVaGr00__} and organic dye aggregates~\cite{Ko96__}, another type of energy transfer has received more recent interest, namely the propagation of internal excitation due to electromagnetic interactions. Experiments indicate robust, coherent energy transport of this kind in systems as large as photosynthetic light harvesting complexes \cite{engel_fleming:coherence_nature:lee_fleming:coherence_science:Collini_scholes:coherence_science}, opening up the avenue to consider the propagation of a particularly ``fragile'' quantum phenomenon: entanglement~\cite{sarovar_fleming:entanglement,caruso:entang_fmo}. Here we show for the example of a chain of ultra-cold Rydberg atoms that mechanical momentum transfer interlinked with coherent excitation
migration can result in efficient transport of entanglement.

Rydberg atoms have recently received much attention, to a large part due to their strong long-range interactions, with
diverse consequences from dipole-blockade~\cite{lukin:quantuminfo:urban:twoatomblock:gaetan:twoatomblock} and anti-blockade~\cite{cenap:antiblockade,amthor_antiblock} over long range
molecules~\cite{Greene:LongRangeMols,liu:ultra_long_range_2009} to classical motion due to Van-der-Waals interactions~\cite{amthor:vanderwaals}. In contrast
to the latter, resonant dipole-dipole interactions \cite{anderson:resonant_dipole,li_gallagher:dipdipexcit,noordam:interactions,cenap:motion} intimately link motion and excitation transport.
Within an essential state picture, where only two electronic Rydberg states per
atom, labelled $\ket{a}$ and $\ket{b}$, are taken into account, the transfer of excitation can be adequately
described by using the exciton theory of Frenkel~\cite{AmVaGr00__}.
For a pair of atoms separated by a distance $R$, dipole-dipole interactions have a Hamiltonian with structure 
$H={V}(R)(\ket{ab}\bra{ba} + \ket{ba}\bra{ab})$, where ${V}(R)$ scales like $R^{-3}$. It describes electronic excitation
transfer, since a transition of the first atom from $\ket{a}$ to $\ket{b}$ is accompanied by the reverse transition of the second atom. 
To see how this also induces mechanical forces, we consider a superposition
of two-atom states like $\ket{ab}\pm \ket{ba}$, which are excitonic
eigenstates of $H$ with eigenvalues $\pm V(R)$ that parametrically depend on
the distance $R$. These provide adiabatic potentials for the nuclear motion. 
The character of the motion (eg.~fully repulsive or fully attractive) depends on the exciton state~\cite{cenap:motion}. Adiabatic motion of atoms in a longer chain preserves the exciton character. Since the exciton state for more than two atoms depends on the atomic positions, excitation transport and motion become interlinked. 

In detail, we study the effect of resonant dipole-dipole interactions on a
regular linear chain of Rydberg atoms. Initially we impose a perturbation in the distances between the atoms by placing two atoms close together, with a localized exciton state built on this diatomic proximity, choosen repulsively. We demonstrate a strong correlation between the resulting exciton dynamics and the
motion of the atoms: The combined pulse of atomic displacements and localized electronic excitation
propagates adiabatically through the chain in a manner reminiscent of Newton's cradle. 

We treat this complex many-body problem using a mixed quantum-classical
approach (Tullys surface hopping method \cite{tully:hopping2,tully:hopping}). It allows us to determine the dynamics of the atomic wave-packets together with the electronic excitation transport in order to quantify the entanglement in time. For short chains, where a full quantum mechanical treatment is possible, the numerically exact solution is in perfect agreement with the quantum-classical result.

%%%%%%%%%%%%%%%%%%%%%%%%%%%%%%%%%%%
%%%%%%%%%%%%%%%%%%%%%%%%%%%%%%%%%%%

\begin{figure}[bt]
\psfig{file=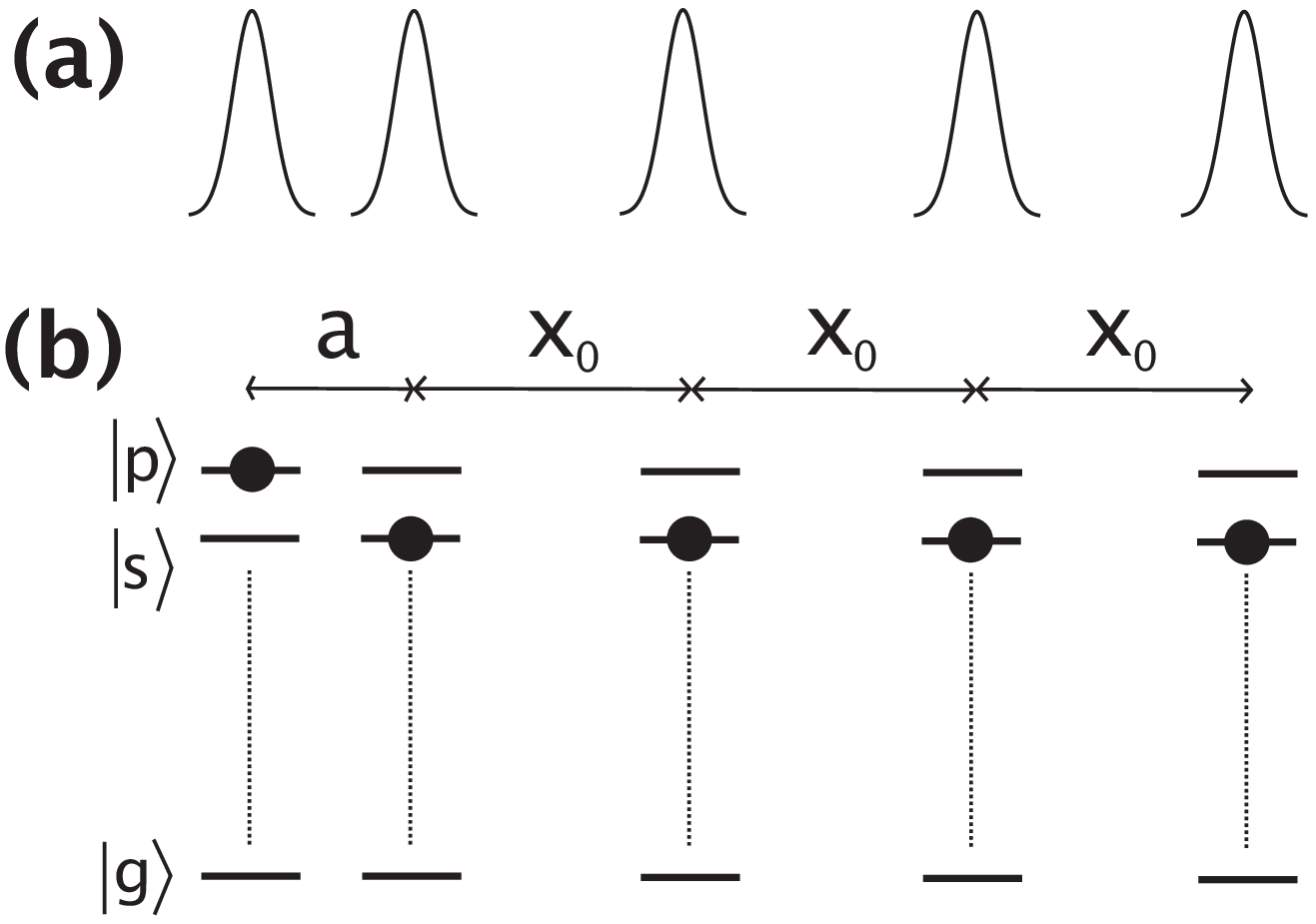,width=0.49\columnwidth}
\psfig{file=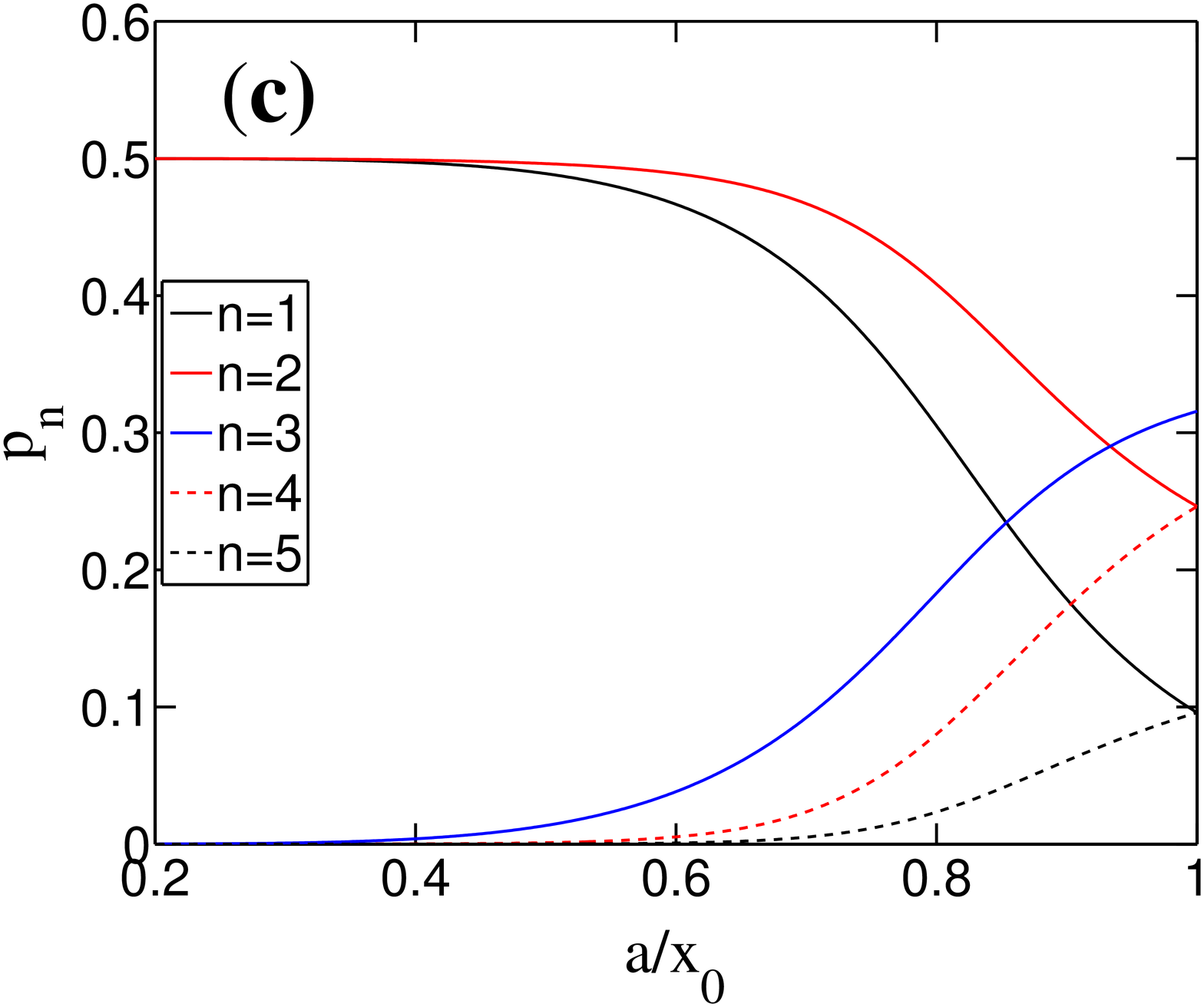,width=0.49\columnwidth}
\caption{\label{fig:sketch}(color-online) (a) Sketch of the initial total density distribution of $N=5$ Rydberg atoms. (b) Visualization of the electronic state $\ket{\pi_1}$. (c) Trapping of the electronic excitation in the repulsive exciton state by a perturbation of the regular chain. Shown are the populations $p_{n}=|\braket{\pi_n}{\sub{\varphi}{rep}(\bv{R})}|^2$ as a function of $a/x_{0}$.}
\end{figure}

We study a linear chain of $N$ identical atoms with mass $M$ and denote by $\Pos_n$
the position of the $n$th atom (nuclear coordinates).
All but one of the $N$ atoms shall be in a Rydberg state $|\nu s\rangle$, i.e.~with principal quantum number $\nu$ and angular momentum $l=0$. Just a single atom is in the state $|\nu p\rangle$, i.e.~has angular momentum $l=1$.
The latter will be called the ``excited'' state hereafter. It can migrate along the chain by means of dipole-dipole interactions~\cite{noordam:interactions}, which conserve the number of excitations. We can restrict ourselves to the single-excitation Hilbert space, whose electronic part is spanned by $\ket{\pi_n}\equiv \ket{s\cdots p\cdots s}$, see \frefp{fig:sketch}{b}, since for the scenario we consider, transitions to other states are negligible. The distance $\PosAbs_{nm}\equiv | \Pos_m-\Pos_n|$ between the atoms $n$ and $m$ 
is so large that the overlap of their electronic wave functions can be neglected. The total Hamiltonian of the system is
\begin{equation}
\label{Ham_Tot}
\HamTot(\bv{R})=-\sum_{n=1}^N\frac{\hbar^2}{2 M} \nabla^2_{\Pos_n} + H^{\rm el}(\bv{R}),
\end{equation}
where $\bv{R}=(\Pos_1,\dots,\Pos_N)^T$ is the vector of nuclear positions. 
The electronic Hamiltonian
\begin{equation}
\label{H_el}
\HamEl(\bv{R})=\sum_{nm}\Vdd_{nm}(\PosAbs_{nm})\ket{\pi_n}\bra{\pi_m}
\end{equation}
contains the dipole-dipole coupling between atoms $n$ and $m$. We consider the case with all atoms in $m_{l}=0$ azimuthal angular momentum states, such that $\Vdd_{nm}(\Pos_{nm})=-\mu^2/\PosAbs_{nm}^3$ \emph{without angular dependence} \cite{noordam:interactions}. 

Our numerical calculations use an atomic mass $M=11000\ \rm{ au}$ (which is roughly the mass of Lithium) and a transition dipole moment $\mu=1000\ \rm{ au}$, corresponding to transitions between $s$ and $p$ states with $n\approx 30 \dots 40$. 

The full many-body problem posed by the Hamiltonian \bref{Ham_Tot} becomes quickly intractable as the number of atoms $N$ is increased. For small $N$ however, it is no problem to directly solve the equation of motion. Expanding the full wave function in electronic (diabatic) states according to $\ket{\Psi(\bv{R})}=\sum_{n=1}^{N} \phi_{n}(\bv{R}) \ket{\pi_n}$, we arrive at the Schr{\"o}dinger equation (in atomic units)
\begin{align}
\label{fullSE}
i \dot{\phi}_{n} (\bv{R})&=\sum_{m=1}^N\left[-\frac{ \nabla^2_{\Pos_m} }{2 M}  \phi_{n}(\bv{R}) + \Vdd_{nm}(\Pos_{nm})\phi_{m}(\bv{R}) \right].
\end{align}
In order to validate the semi-classical method presented below, which in turn will be faithfully used for longer chains, we solve \eref{fullSE} for $N=3$.
In our figures we will not show the full $N$-dimensional nuclear wave function but focus on the more intuitive total atomic density, which is given by $
n(R)= \sum_{j=1}^N\sum_{m=1}^N \int d^{N-1}\bv{R}_{\{j\}} |\phi_{m}(\bv{R})|^{2}$. 
Here $\int d^{N-1}\bv{R}_{\{j\}}$ denotes integration over all but the $j-th$ nuclear coordinate.
The density $n(R)$ gives the probability to find an atom at position $R$.

The \emph{diabatic} representation of the wave function allows a
straight forward propagation for short chains. For longer chains and for the interpretation of the results, the \emph{adiabatic} representation $\ket{\Psi(\bv{R})}=\sum_{n=1}^{N} \tilde{\phi}_{n}(\bv{R})\ket{\varphi_n(\bv{R})}$ is helpful. Here the adiabatic basis $\ket{\varphi_{n}}$ is defined via $H^{\rm el}(\bv{R} )\ket{\varphi_{n}(\bv{R})}=U_{n}(\bv{R} )\ket{\varphi_{n}(\bv{R})}$. 
For each $\bv{R}$ there are $N$ excitonic eigenstates $\ket{\varphi_{n}(\bv{R})}$ labeled by the index $n$. The corresponding eigenenergies $U_{n}(\bv{R} )$ define the adiabatic potential surfaces.
The two representations are related by $\tilde{\phi}_{n}(\bv{R}) = \sum_{m} \braket{\varphi_n(\bv{R})}{\pi_m}\phi_{m}(\bv{R})$. 

%%%%%%%%%%%%%%%%%%%%%%%%%%%%%%%%%%%
%%%%%%%%%%%%%%%%%%%%%%%%%%%%%%%%%%%
For long chains, we solve the time-dependent  Schr\"odinger equation with Hamiltonian (\ref{Ham_Tot}) using 
a mixed quantum/classical method, Tully's surface hopping algorithm~\cite{tully:hopping2,tully:hopping}. In this approach an ensemble of trajectories is propagated, and each trajectory moves classically on a \emph{single} adiabatic surface $U_{m}(\bv{R})$, except for the possibility of instantaneous switches among the adiabatic states.

The equations of motion read
\begin{align}
\label{TullysEOM_elec}
i \pdiff{}{t} \tilde{c}_{k} &=U_{k}(\bv{R})\tilde{c}_{k} - i \sum_{q=1}^{N} \dot{\bv{R}} \cdot \bv{d}_{kq} \tilde{c}_{q}, 
\\
\label{TullysEOM_classic}
M \ddot{\bv{R}}&=-\bv{\nabla}_{\bv{R}}\langle \varphi_m(\bv{R}) | \HamEl(\bv{R}) | \varphi_m(\bv{R}) \rangle.
\end{align}

The $N$ complex amplitudes $\tilde{c}_{k}$ define the electronic state via $\ket{\Psi(\bv{R},t)}=\sum_{n=1}^{N}\tilde{c}_{n}(t) \ket{\varphi_n(\bv{R},t)}$ and the $\bv{d}_{kq}$ are non-adiabatic coupling vectors. Besides their appearance in \eref{TullysEOM_elec}, they also control the likelihood of stochastic jumps from the current surface $m$ to another surface $m'$ in \eref{TullysEOM_classic}, which is proportional to $| \dot{\bv{R}} \cdot \bv{d}_{m'm}|$. Further details about this scheme can be found in Refs.~\cite{cenap:motion,tully:hopping,tully:hopping:veloadjust}. The underlying quantum-classical correspondence is discussed in Refs.~\cite{tully:derivation,delos:semiclass}. We randomize the initial classical positions and velocities for the trajectories according to the Wigner distribution of the initial state, described in \eref{psi_tot_ini}. This is essential for a correct description.

%%%%%%%%%%%%%%%%%%%%%%%%%%%%%%%%%%%
Initially we assume that the Rydberg atoms are arranged in a straight line.
The distance between the first two atoms is denoted by $a$ and assumed shorter than the equal distances ($x_0$) between the other atoms, as sketched in \fref{fig:sketch}. 

For fixed classical positions, one of the $N$ eigenstates of the electronic Hamiltonian \bref{H_el} leads to a situation where initially all atoms repel each other \cite{cenap:motion}. In the following we will focus on this state, which we label with ``rep''. Since the dipole-dipole interaction between the first two atoms is much stronger than between all others, the excitation in this repulsive state is mainly localized on these two atoms. For $a\ll x_0$ this initial state can be approximately written as $(\ket{\pi_1}-\ket{\pi_2})/\sqrt{2}$.
In \frefp{fig:sketch}{c} the electronic population on the various atoms is shown as a function of $a/x_0$. 
For our simulations, we have taken $x_0=5 \mu$m and $a=2 \mu$m, i.e.\ $a/x_0=0.4$. Then the dipole-dipole interaction between the last two atoms is about 5 times larger than for the rest of the chain and more than 90\% of the excitation is localised on the first two atoms. From the point-like (classical) nuclear positions, we now move to a quantum nuclear wave function.

The spatial wave function of each atom is assumed Gaussian with a standard deviation $\sigma_{0}$. Hence, we take the complete initial wave function (i.e.\ containing nuclear and
excitonic degrees of freedom) as
\begin{align}
\label{psi_tot_ini}
\ket{\Psi(t=0)}&=\ket{\varphi_{\rm rep}(\bv{R})}  \prod_{n=1}^N \phi_{\rm G}(\Pos_n),
\CR
\phi_{\rm G}(\Pos_n)&={\cal N}\exp{(-[\Pos_n-\Pos_{0n}]^2/2 \sigma_0^2 )},
\end{align}
where $\Pos_{0n}$ is the center of mass of the $n$-th Gaussian and ${\cal N}$ a normalization factor.

\begin{figure}[bt]
\centerline{\psfig{file=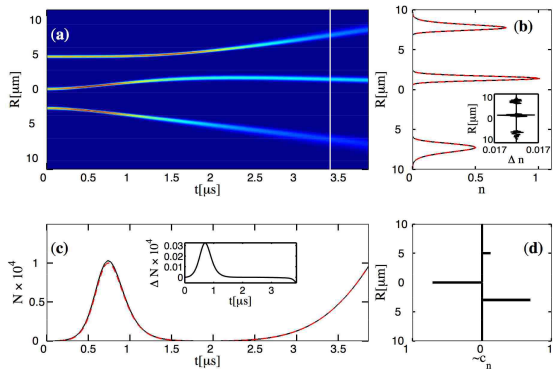,width= \columnwidth}}
\caption{\label{fig:Tully_vs_QM} (color online)  Nuclear dynamics in the
 case $N=3$.   The time evolution of the total atomic
 density $n(x,t)$ (a) is shown together with a comparison of Tully's
 surface hopping calculations (black solid) with the full quantum evolution
 (red dashed) in the other panels. (b) Spatial slice $n(x,t_{0})$, with $t_{0}$ as indicated
by the first vertical white lines in (a). Arbitrary units. (c) Relative population  $\sub{n}{2}=\int d \bv{R} |\tilde{\phi}_{2}(\bv{R})|^{2}$
($\sub{n}{2}=|\sub{\tilde{c}}{2}|^{2}$ in the Tully algorithm) on the adiabatic
surface (index $2$) that is energetically nearest to the initial repulsive one. This is a measure of
the propensity of non-adiabatic transitions. The deviation of curves in (b,c) is shown magnified in the insets. (d) Initial repulsive state.
}
\end{figure}

%%%%%%%%%%%%%%%%%%%%%%%%%%%%%%%%%%%
%%%%%%%%%%%%%%%%%%%%%%%%%%%%%%%%%%%
To confirm the applicability of the quantum-classical numerical treatment, we consider the smallest non-trivial chain $N=3$. Figure \ref{fig:Tully_vs_QM} shows the quantum mechanical probability to find an atom at a certain position, in perfect agreement with the corresponding graph obtained with the quantum-classical hybrid approach. 

\begin{figure}[bt]
\centerline{\psfig{file=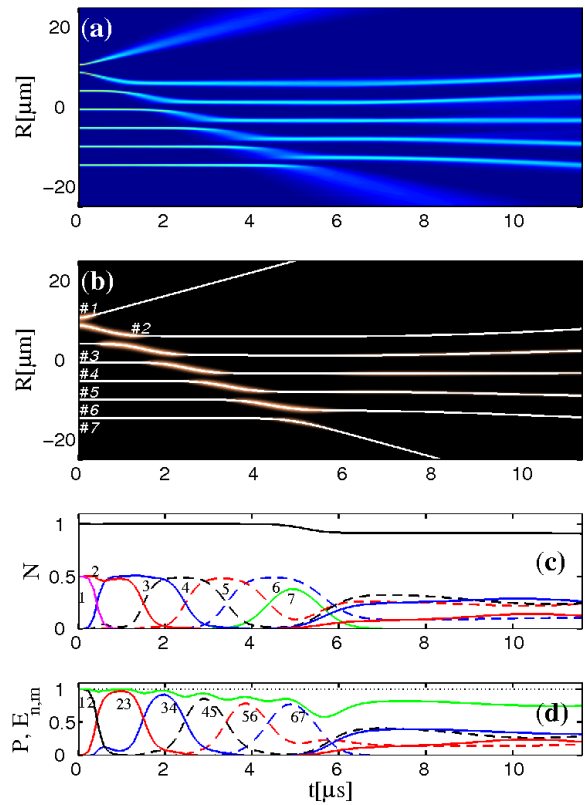,width= \columnwidth}}
\caption{\label{fig:longchain} (color online) Dynamics of atomic motion and excitation transfer. (a) Total atomic density averaged over $10^5$ realizations. We actually plot $\sqrt{n}$. (b) Mean trajectories of the individual atoms (white) and electronic excitation probabilities (diabatic populations) $|c_{m}|^2$, $c_{m}=\sum_{m} O_{nm}^{T}\tilde{c}_{m}$. The latter are encoded as the width of the copper shading surrounding each trajectory. (c) Population on the adiabatic surface $rep$ (black) and individual diabatic populations. (d) Purity $P=$Tr$[\hat{\sigma}^{2}]$ of the reduced electronic density matrix $\hat{\sigma}$ (green) and bipartite entanglement $E_{n,n+1}$ for neighboring atoms as defined in the text. The dotted line indicates $1$.}
\end{figure}

The excellent agreement between the two disparate methods for $N=3$ gives confidence that Tully's surface hopping produces reliable results also for longer chains, such as $N=7$, which we consider now. The corresponding atomic motion and excitation transfer, when starting in the exciton state with highest energy are shown in \fref{fig:longchain}. 
Let us first consider the atomic motion. As expected, initially the two excited atoms strongly repel each other. 
When atom {\atomA} has approached atom {\atomB}, the main repulsion is now between those two, causing atom {\atomA} to slow down and atom {\atomB} to accelerate.
In this way the initial momentum is transferred through the chain to atom 7, realizing a microscopic version of Newton's cradle. 

To lay the basis for the treatment of entanglement dynamics, we next discuss the excitation transfer, shown in \frefp{fig:longchain}{c}, which is strongly
coupled to the atomic motion as can be seen in \frefp{fig:longchain}{b}. The excitation gets transferred remaining always localized
on the two instantaneously nearest atoms, in accordance with the structure of
exciton eigenstates outlined in \cite{cenap:motion}. After $5.5$ $\mu$s the
momentum transferred through the chain kicks out the last atom, and a
well defined close proximity pair no longer exists. The exciton state then
assumes the shape for an equidistant chain, de-localized over the entire chain
(consisting of the remaining $N$-2 atoms), which subsequently slowly spreads out. From the occupation of the initially populated repulsive adiabatic state, which is also shown in \frefp{fig:longchain}{b}, we can deduce that the evolution is largely adiabatic. Non-adiabatic effects occur only in a small interval around $t=5.5\mu$s, when the last atom leaves the chain.

So far we have described transfer of momentum and kinetic energy through the Rydberg chain, both of which would also occur in a classical Newton's cradle. However, the microscopic excitation migration leads also to a transport of entanglement, which has no classical equivalent. In the spirit of Newton's cradle we focus on electronic entanglement between two atoms, which we will quantify with the bipartite \emph{entanglement of formation} \cite{wootters:mixed,hill:wootters:qbits}. To this end we need the reduced density matrix $\hat{\sigma}=\sum_{n,m}\sigma_{nm} \ket{\pi_{n}}\bra{\pi_{m}}$, describing the electronic state of the system, after tracing over the atomic positions. The matrix elements $\sigma_{nm}$ are given by  $\sigma_{nm}=\int d^N \bv{R}\:\:  \phi^*_{n}(\bv{R}) \phi_{m}(\bv{R})$ for the full quantum calculations and $\sigma_{nm}=\overline{c_{n}^* c_{m}}$ for the surface hopping method. In the latter case $\overline{\cdots}$
denotes the trajectory average and $c_{n}=\sum_{m} O_{nm}^T \tilde{c}_{m}$ are the coefficients in the diabatic basis.
From $\hat{\sigma}$ we then construct the binary reduced electronic density
matrix of atoms $a$ and $b$ by $\hat{\beta}_{ab}={\mbox{Tr}}^{\{a,b\}}\big[\hat{\sigma}\big]$. The symbol ${\mbox{Tr}}^{\{a,b\}}\big[\cdots\big]$ denotes the trace over the
electronic states for all atoms other than $a$, $b$. The remaining reduced subspace of atoms $a$ and $b$  is spanned by $\ket{pp}$,
$\ket{ps}$,  $\ket{sp}$,  $\ket{ss}$. Due to the structure of the $\ket{\pi_n}$, the only non-vanishing matrix elements of $\hat{\beta}_{ab}$ are $\bra{ps} \hat{\beta}_{ab} \ket{ps}=\sigma_{aa}$, $\bra{sp} \hat{\beta}_{ab} \ket{sp}=\sigma_{bb}$, $\bra{ps} \hat{\beta}_{ab} \ket{sp}=\bra{sp} \hat{\beta}_{ab} \ket{ps}^*=\sigma_{ba}$ and $\bra{ss}\hat{\beta}_{ab} \ket{ss}=\sum_{c\neq \{a,b\}} \sigma_{cc}$. From the matrix $\hat{\beta}$ we can derive the concurrence $C_{ab}=2|\sigma_{ab}|$ as outlined in \cite{wootters:mixed}. The concurrence is already a measure of entanglement with $0\leq C_{ab} \leq 1$, from which we finally obtain the entanglement of formation $0\leq E_{ab}(C_{ab})\leq 1$ as described in \cite{wootters:mixed,hill:wootters:qbits}.

As can be seen in \frefp{fig:longchain}{d}, the initially
perfect entanglement between atom 1 and 2 is transported through the chain with
only minor losses up to the point where the final atom leaves the chain around $t=5.5\mu$s \cite{footnote:entanglecheck}. At that moment, the exciton state de-localizes over the entire chain. Entanglement is then shared among all remaining atoms, with a resulting drop of
\emph{bipartite} entanglement.

%%%%%%%%%%%%%%%%%%%%%%%%%%%%%%%%%%%
%%%%%%%%%%%%%%%%%%%%%%%%%%%%%%%%%%%
In summary, for an aggregate of Rydberg atoms in form of a linear chain, we have identified a dynamical mode that links the motion of the atoms intimately with the coherent propagation of a single
electronic excitation along the chain. Adiabatic transport ensures that the excitation remains spatially localized near a diatomic proximity passing through the chain. Akin to the transfer of the almost macroscopic quantities, energy and momentum, in Newton's cradle, the mode transports localized coherent excitation and electronic entanglement along the chain. For the underlying quantum many-body problem we have demonstrated the applicability of Tully's surface hopping method~\cite{tully:hopping2,tully:hopping} by comparison with exact calculations.

%%%%%%%%%%%%%%%%%%%%%%%%%%%%%%%%%%%

%%%%%%%%%%%%%%%%%%%%%%%%%%%%%%%%%%%

%%%%%%%%%%%%%%%%%%%%%%%%%%%%%%%%%%%%%%%%%%%%


\begin{thebibliography}{29}
\expandafter\ifx\csname natexlab\endcsname\relax\def\natexlab#1{#1}\fi
\expandafter\ifx\csname bibnamefont\endcsname\relax
  \def\bibnamefont#1{#1}\fi
\expandafter\ifx\csname bibfnamefont\endcsname\relax
  \def\bibfnamefont#1{#1}\fi
\expandafter\ifx\csname citenamefont\endcsname\relax
  \def\citenamefont#1{#1}\fi
\expandafter\ifx\csname url\endcsname\relax
  \def\url#1{\texttt{#1}}\fi
\expandafter\ifx\csname urlprefix\endcsname\relax\def\urlprefix{URL }\fi
\providecommand{\bibinfo}[2]{#2}
\providecommand{\eprint}[2][]{\url{#2}}

\bibitem[{\citenamefont{Herrmann and Schm{\"a}lzle}(1981)}]{herrmann:cradle}
\bibinfo{author}{\bibfnamefont{F.}~\bibnamefont{Herrmann}} \bibnamefont{and}
  \bibinfo{author}{\bibfnamefont{P.}~\bibnamefont{Schm{\"a}lzle}},
  \bibinfo{journal}{Am. J. Phys.} \textbf{\bibinfo{volume}{49}},
  \bibinfo{pages}{761} (\bibinfo{year}{1981}).

\bibitem[{\citenamefont{Kinoshita et~al.}(2006)\citenamefont{Kinoshita, Wenger,
  and Weiss}}]{kinoshita:cradle}
\bibinfo{author}{\bibfnamefont{T.}~\bibnamefont{Kinoshita}},
  \bibinfo{author}{\bibfnamefont{T.}~\bibnamefont{Wenger}}, \bibnamefont{and}
  \bibinfo{author}{\bibfnamefont{D.~S.} \bibnamefont{Weiss}},
  \bibinfo{journal}{Nature} \textbf{\bibinfo{volume}{440}},
  \bibinfo{pages}{900} (\bibinfo{year}{2006}).

\bibitem[{\citenamefont{van Amerongen et~al.}(2000)\citenamefont{van Amerongen,
  Valkunas, and van Grondelle}}]{AmVaGr00__}
\bibinfo{author}{\bibfnamefont{H.}~\bibnamefont{van Amerongen}},
  \bibinfo{author}{\bibfnamefont{L.}~\bibnamefont{Valkunas}}, \bibnamefont{and}
  \bibinfo{author}{\bibfnamefont{R.}~\bibnamefont{van Grondelle}},
  \emph{\bibinfo{title}{{Photosynthetic Excitons}}} (\bibinfo{publisher}{World
  Scientific, Singapore}, \bibinfo{year}{2000}).

\bibitem[{\citenamefont{Kobayashi}(1996)}]{Ko96__}
\bibinfo{editor}{\bibfnamefont{T.}~\bibnamefont{Kobayashi}}, ed.,
  \emph{\bibinfo{title}{{J-Aggregates}}} (\bibinfo{publisher}{World
  Scientific}, \bibinfo{year}{1996}).

\bibitem[{\citenamefont{Engel et~al.}(2007)\citenamefont{Engel, Calhoun, Read,
  {T.-K. Ahn}, Man{\v c}al, {Y.-C. Cheng}, Blankenship, and
  Fleming}}]{engel_fleming:coherence_nature:lee_fleming:coherence_science:Collini_scholes:coherence_science}
\bibinfo{author}{\bibfnamefont{G.~S.} \bibnamefont{Engel} \bibnamefont{et~al.} }, \bibinfo{journal}{Nature}
  \textbf{\bibinfo{volume}{446}}, \bibinfo{pages}{782} (\bibinfo{year}{2007}). 
  \bibinfo{author}{\bibfnamefont{H.}~\bibnamefont{Lee} \bibnamefont{et~al.}},
  \bibinfo{journal}{Science} \textbf{\bibinfo{volume}{316}},
  \bibinfo{pages}{1462} (\bibinfo{year}{2007}). \bibinfo{author}{\bibfnamefont{E.}~\bibnamefont{Collini}} \bibnamefont{and}
  \bibinfo{author}{\bibfnamefont{G.~D.} \bibnamefont{Scholes}},
  \bibinfo{journal}{Science} \textbf{\bibinfo{volume}{323}},
  \bibinfo{pages}{369} (\bibinfo{year}{2009}).

\bibitem[{\citenamefont{Sarovar et~al.}(2010)\citenamefont{Sarovar, Ishizaki,
  Fleming, and Whaley}}]{sarovar_fleming:entanglement}
\bibinfo{author}{\bibfnamefont{M.}~\bibnamefont{Sarovar}},
  \bibinfo{author}{\bibfnamefont{A.}~\bibnamefont{Ishizaki}},
  \bibinfo{author}{\bibfnamefont{G.~R.} \bibnamefont{Fleming}},
  \bibnamefont{and} \bibinfo{author}{\bibfnamefont{K.~B.}
  \bibnamefont{Whaley}}, \bibinfo{journal}{Nature Physics}
  \textbf{\bibinfo{volume}{6}}, \bibinfo{pages}{462} (\bibinfo{year}{2010}).

\bibitem[{\citenamefont{Caruso et~al.}(2009)\citenamefont{Caruso, Chin, Datta,
  Huelga, and Plenio}}]{caruso:entang_fmo}
\bibinfo{author}{\bibfnamefont{F.}~\bibnamefont{Caruso}},
  \bibinfo{author}{\bibfnamefont{A.~W.} \bibnamefont{Chin}},
  \bibinfo{author}{\bibfnamefont{A.}~\bibnamefont{Datta}},
  \bibinfo{author}{\bibfnamefont{S.~F.} \bibnamefont{Huelga}},
  \bibnamefont{and} \bibinfo{author}{\bibfnamefont{M.~B.}
  \bibnamefont{Plenio}}, \bibinfo{journal}{J. Chem. Phys.}
  \textbf{\bibinfo{volume}{131}}, \bibinfo{pages}{105106}
  (\bibinfo{year}{2009}).

\bibitem[{\citenamefont{Lukin et~al.}(2001)\citenamefont{Lukin, Fleischhauer,
  {C\^ot\'e}, Duan, Jaksch, Cirac, and Zoller}}]{lukin:quantuminfo:urban:twoatomblock:gaetan:twoatomblock}
\bibinfo{author}{\bibfnamefont{M.~D.} \bibnamefont{Lukin} \bibnamefont{et~al.}},
  \bibinfo{journal}{Phys. Rev. Lett.} \textbf{\bibinfo{volume}{87}},
  \bibinfo{pages}{037901} (\bibinfo{year}{2001}). \bibinfo{author}{\bibfnamefont{E.}~\bibnamefont{Urban} \bibnamefont{et~al.}},
  \bibinfo{journal}{Nature Physics} \textbf{\bibinfo{volume}{5}},
  \bibinfo{pages}{110} (\bibinfo{year}{2009}). \bibinfo{author}{\bibfnamefont{A.}~\bibnamefont{Ga{\"e}tan} \bibnamefont{et~al.}},
  \bibinfo{journal}{Nature Physics} \textbf{\bibinfo{volume}{5}},
  \bibinfo{pages}{115} (\bibinfo{year}{2009}).

\bibitem[{\citenamefont{Ates et~al.}(2007)\citenamefont{Ates, Pohl, Pattard,
  and Rost}}]{cenap:antiblockade}
\bibinfo{author}{\bibfnamefont{C.}~\bibnamefont{Ates}},
  \bibinfo{author}{\bibfnamefont{T.}~\bibnamefont{Pohl}},
  \bibinfo{author}{\bibfnamefont{T.}~\bibnamefont{Pattard}}, \bibnamefont{and}
  \bibinfo{author}{\bibfnamefont{J.~M.} \bibnamefont{Rost}},
  \bibinfo{journal}{Phys. Rev. Lett.} \textbf{\bibinfo{volume}{98}},
  \bibinfo{pages}{023002} (\bibinfo{year}{2007}).

\bibitem[{\citenamefont{Amthor et~al.}(2010)\citenamefont{Amthor, Giese,
  Hofmann, and Weidem\"uller}}]{amthor_antiblock}
\bibinfo{author}{\bibfnamefont{T.}~\bibnamefont{Amthor}},
  \bibinfo{author}{\bibfnamefont{C.}~\bibnamefont{Giese}},
  \bibinfo{author}{\bibfnamefont{C.~S.} \bibnamefont{Hofmann}},
  \bibnamefont{and}
  \bibinfo{author}{\bibfnamefont{M.}~\bibnamefont{Weidem\"uller}},
  \bibinfo{journal}{Phys. Rev. Lett.} \textbf{\bibinfo{volume}{104}},
  \bibinfo{pages}{013001} (\bibinfo{year}{2010}).

\bibitem[{\citenamefont{Greene et~al.}(2000)\citenamefont{Greene, Dickinson,
  and Sadeghpour}}]{Greene:LongRangeMols}
\bibinfo{author}{\bibfnamefont{C.~H.} \bibnamefont{Greene}},
  \bibinfo{author}{\bibfnamefont{A.~S.} \bibnamefont{Dickinson}},
  \bibnamefont{and} \bibinfo{author}{\bibfnamefont{H.~R.}
  \bibnamefont{Sadeghpour}}, \bibinfo{journal}{Phys. Rev. Lett.}
  \textbf{\bibinfo{volume}{85}}, \bibinfo{pages}{2458} (\bibinfo{year}{2000}).

\bibitem[{\citenamefont{Liu et~al.}(2009)\citenamefont{Liu, Stanojevic, and
  Rost}}]{liu:ultra_long_range_2009}
\bibinfo{author}{\bibfnamefont{I.~C.~H.} \bibnamefont{Liu}},
  \bibinfo{author}{\bibfnamefont{J.}~\bibnamefont{Stanojevic}},
  \bibnamefont{and} \bibinfo{author}{\bibfnamefont{J.~M.} \bibnamefont{Rost}},
  \bibinfo{journal}{Phys. Rev. Lett.} \textbf{\bibinfo{volume}{102}},
  \bibinfo{pages}{173001} (\bibinfo{year}{2009}).

\bibitem[{\citenamefont{Amthor et~al.}(2007)\citenamefont{Amthor,
  {Reetz-Lamour}, Giese, and Weidem{\"u}ller}}]{amthor:vanderwaals}
\bibinfo{author}{\bibfnamefont{T.}~\bibnamefont{Amthor}},
  \bibinfo{author}{\bibfnamefont{M.}~\bibnamefont{{Reetz-Lamour}}},
  \bibinfo{author}{\bibfnamefont{C.}~\bibnamefont{Giese}}, \bibnamefont{and}
  \bibinfo{author}{\bibfnamefont{M.}~\bibnamefont{Weidem{\"u}ller}},
  \bibinfo{journal}{Phys. Rev. A} \textbf{\bibinfo{volume}{76}},
  \bibinfo{pages}{054702} (\bibinfo{year}{2007}).

\bibitem[{\citenamefont{Anderson et~al.}(1998)\citenamefont{Anderson, Veale,
  and Gallagher}}]{anderson:resonant_dipole}
\bibinfo{author}{\bibfnamefont{W.~R.} \bibnamefont{Anderson}},
  \bibinfo{author}{\bibfnamefont{J.~R.} \bibnamefont{Veale}}, \bibnamefont{and}
  \bibinfo{author}{\bibfnamefont{T.~F.} \bibnamefont{Gallagher}},
  \bibinfo{journal}{Phys. Rev. Lett.} \textbf{\bibinfo{volume}{80}},
  \bibinfo{pages}{249} (\bibinfo{year}{1998}).

\bibitem[{\citenamefont{Li et~al.}(2005)\citenamefont{Li, Tanner, and
  Gallagher}}]{li_gallagher:dipdipexcit}
\bibinfo{author}{\bibfnamefont{W.}~\bibnamefont{Li}},
  \bibinfo{author}{\bibfnamefont{P.~J.} \bibnamefont{Tanner}},
  \bibnamefont{and} \bibinfo{author}{\bibfnamefont{T.~F.}
  \bibnamefont{Gallagher}}, \bibinfo{journal}{Phys. Rev. Lett.}
  \textbf{\bibinfo{volume}{94}}, \bibinfo{pages}{173001}
  (\bibinfo{year}{2005}).

\bibitem[{\citenamefont{Robicheaux et~al.}(2004)\citenamefont{Robicheaux,
  Hernandez, Topcu, and Noordam}}]{noordam:interactions}
\bibinfo{author}{\bibfnamefont{F.}~\bibnamefont{Robicheaux}},
  \bibinfo{author}{\bibfnamefont{J.~V.} \bibnamefont{Hernandez}},
  \bibinfo{author}{\bibfnamefont{T.}~\bibnamefont{Topcu}}, \bibnamefont{and}
  \bibinfo{author}{\bibfnamefont{L.~D.} \bibnamefont{Noordam}},
  \bibinfo{journal}{Phys. Rev. A} \textbf{\bibinfo{volume}{70}},
  \bibinfo{pages}{042703} (\bibinfo{year}{2004}).

\bibitem[{\citenamefont{Ates et~al.}(2008)\citenamefont{Ates, Eisfeld, and
  Rost}}]{cenap:motion}
\bibinfo{author}{\bibfnamefont{C.}~\bibnamefont{Ates}},
  \bibinfo{author}{\bibfnamefont{A.}~\bibnamefont{Eisfeld}}, \bibnamefont{and}
  \bibinfo{author}{\bibfnamefont{J.~M.} \bibnamefont{Rost}},
  \bibinfo{journal}{New J. Phys.} \textbf{\bibinfo{volume}{10}},
  \bibinfo{pages}{045030} (\bibinfo{year}{2008}).

\bibitem[{\citenamefont{Tully and Preston}(1971)}]{tully:hopping2}
\bibinfo{author}{\bibfnamefont{J.~C.} \bibnamefont{Tully}} \bibnamefont{and}
  \bibinfo{author}{\bibfnamefont{R.~K.} \bibnamefont{Preston}},
  \bibinfo{journal}{J. Chem. Phys.} \textbf{\bibinfo{volume}{55}},
  \bibinfo{pages}{562} (\bibinfo{year}{1971}).

\bibitem[{\citenamefont{Tully}(1990)}]{tully:hopping}
\bibinfo{author}{\bibfnamefont{J.~C.} \bibnamefont{Tully}},
  \bibinfo{journal}{J. Chem. Phys.} \textbf{\bibinfo{volume}{93}},
  \bibinfo{pages}{1061} (\bibinfo{year}{1990}).

\bibitem[{\citenamefont{{Hammes-Schiffer} and
  Tully}(1994)}]{tully:hopping:veloadjust}
\bibinfo{author}{\bibfnamefont{S.}~\bibnamefont{{Hammes-Schiffer}}}
  \bibnamefont{and} \bibinfo{author}{\bibfnamefont{J.~C.} \bibnamefont{Tully}},
  \bibinfo{journal}{J. Chem. Phys.} \textbf{\bibinfo{volume}{101}},
  \bibinfo{pages}{4657} (\bibinfo{year}{1994}).

\bibitem[{\citenamefont{Tully}(1998)}]{tully:derivation}
\bibinfo{author}{\bibfnamefont{J.~C.} \bibnamefont{Tully}},
  \bibinfo{journal}{Faraday Discuss.} \textbf{\bibinfo{volume}{110}},
  \bibinfo{pages}{407} (\bibinfo{year}{1998}).

\bibitem[{\citenamefont{Delos et~al.}(1972)\citenamefont{Delos, Thorson, and
  Knudson}}]{delos:semiclass}
\bibinfo{author}{\bibfnamefont{J.~B.} \bibnamefont{Delos}},
  \bibinfo{author}{\bibfnamefont{W.~R.} \bibnamefont{Thorson}},
  \bibnamefont{and} \bibinfo{author}{\bibfnamefont{S.~K.}
  \bibnamefont{Knudson}}, \bibinfo{journal}{Phys. Rev. A}
  \textbf{\bibinfo{volume}{6}}, \bibinfo{pages}{709} (\bibinfo{year}{1972}).

\bibitem[{\citenamefont{Wootters}(1998)}]{wootters:mixed}
\bibinfo{author}{\bibfnamefont{W.~K.} \bibnamefont{Wootters}},
  \bibinfo{journal}{Phys. Rev. Lett.} \textbf{\bibinfo{volume}{80}},
  \bibinfo{pages}{2245} (\bibinfo{year}{1998}).

\bibitem[{\citenamefont{Hill and Wootters}(1997)}]{hill:wootters:qbits}
\bibinfo{author}{\bibfnamefont{S.}~\bibnamefont{Hill}} \bibnamefont{and}
  \bibinfo{author}{\bibfnamefont{W.~K.} \bibnamefont{Wootters}},
  \bibinfo{journal}{Phys. Rev. Lett.} \textbf{\bibinfo{volume}{78}},
  \bibinfo{pages}{5022} (\bibinfo{year}{1997}).

\bibitem[{foo()}]{footnote:entanglecheck}
\bibinfo{note}{{We have verified that both methods outlined previously give the
  same entanglement evolution for the case $N=3$.}}

\end{thebibliography}
\end{document}